\documentclass[
 reprint,
 amsmath,amssymb, pra, aps,
 dvipsnames
]{revtex4-1}

\usepackage{graphicx}
\usepackage{dcolumn}
\usepackage{bm}
\usepackage[separate-uncertainty=true, alsoload=synchem]{siunitx}
\usepackage[hidelinks]{hyperref}
\usepackage[utf8]{inputenc}

\usepackage{esvect}
\usepackage{pgf}
\usepackage{libertine}
\usepackage{libertinust1math}
\usepackage{blindtext}

\DeclareSIUnit{\arbitraryunit}{arb.\, u.}
\DeclareSIUnit{\countspersecond}{cps}

\makeatletter
\DeclareRobustCommand{\AA}{%
  \leavevmode
  \vbox{\ialign{##\cr
    \hidewidth\char'27 \hidewidth\cr
    \noalign{\nointerlineskip\kern-1.4ex}
    A\cr
  }}%
}
\makeatother

\begin{document}
\preprint{APS/123-QED}

\title{
Direct Measurement of Quantum Efficiency of Single Photon Emitters in Hexagonal Boron Nitride
}
\author{Niko Nikolay$^{1,2}$}
\author{Noah Mendelson$^3$}
\author{Ersan Özelci$^{1,2}$}
\author{Bernd Sontheimer$^{1,2}$}
\author{Florian Böhm$^{1,2}$}
\author{Günter Kewes$^{1,2}$}
\author{Milos Toth$^3$}
\author{Igor Aharonovich$^3$}
\author{Oliver Benson$^{1,2}$}
\affiliation{
 $^1$ AG Nanooptik, Humboldt-Universität zu Berlin,
 Newtonstraße 15, D-12489 Berlin, Germany \\
 $^2$ IRIS Adlershof, Humboldt-Universität zu Berlin,
 Zum Großen Windkanal 6, 12489 Berlin, Germany\\
 $^3$ School of Mathematical and Physical Sciences, University of Technology Sydney, Ultimo, New South Wales 2007, Australia
}
\date{\today}
\begin{abstract}

Single photon emitters in two-dimensional materials are promising candidates for future generation of quantum photonic technologies. In this work, we experimentally determine the quantum efficiency (QE) of single photon emitters (SPE) in few-layer hexagonal boron nitride (hBN). We employ a metal hemisphere that is attached to the tip of an atomic force microscope to directly measure the lifetime variation of the SPEs as the tip approaches the hBN. This technique enables non-destructive, yet direct and absolute measurement of the QE of SPEs. We find that the emitters exhibit very high QEs approaching \SI{87(7)}{\percent} at wavelengths of $\approx\,$\SI{580}{\nano\meter}, which is amongst the highest QEs recorded for a solid state single photon emitter.

\end{abstract}

\maketitle

\section{Introduction}

Two-dimensional (2D) materials exhibit unique optoelectronic, nanophotonic and quantum effects, that are not possible with their bulk counterparts \cite{Novoselov2016, Mak2016, Urbaszek2019, Toth2019}. Hexagonal Boron Nitride (hBN) is one such material, that has attracted considerable attention due to its ability to host ultra bright single photon emitters (SPEs) that operate at room temperature \cite{Tran2016b, Kianinia2017a, Proscia2018a, Kim2018, jungwirth_temperature_2016, Grosso2017a}. The emitters are point defects (impurity, missing atoms or vacancy complexes), embedded in the lattice of hBN. Recent efforts have focused on understanding the photophysical properties of these defects, with the goal to increase their brightness, stability, and yield \cite{Tawfik2017, Wigger2019, Feldman2019, NgocMyDuong2018, Vogl2018, Xu2018a}.

An important parameter that up to now has been unknown for the family of SPEs in hBN is their quantum efficiency (QE). The QE is an important parameter of any light source to be considered for implementation in practical devices. However, measuring the quantum efficiency of solid state sources is challenging due to the significantly varying surrounding electromagnetic environment. 

In this work, we utilise a recently engineered family of SPEs in a few nm thick hBN films, grown by chemical vapor deposition (CVD). These exhibit less wavelength variability than commercial hBN sources and also provide a flat topography over large area \cite{Mendelson2019, Comtet2019, Lin2017}, which is advantageous for our experiments. To measure the QE, we utilise a method that was pioneered by Drexhage in the 1970s, \cite{Drexhage1970a} who investigated changes in the intrinsic radiative decay rate of europium ions as a function of distance to a silver mirror. The underlying modification of spontaneous emission of the emitters in close proximity to a metal surface is a quantum electrodynamic effect and related to local density of states (LDOS)\cite{Barnes1998, Novotny2006, Schell2014, Beams2013, Anger2006}. Since the intrinsic non-radiative decay rate is not modified by the LDOS, changes in the total decay rate can be attributed to an alteration of the radiative decay rate alone. In this way, radiative and non-radiative decay rate components can be separated by recording the emitter’s excited state lifetime as a function of its distance to a mirror. Hence, this fundamental technique to directly measure the QE from decay rate components has been used for organic dyes, rare earth ions, quantum dots, and color centers in diamond. \cite{Buchler2005,Lunnemann2013,Karaveli2013, Kwadrin2012}

\begin{figure}[htb]
 \includegraphics[width=\columnwidth]{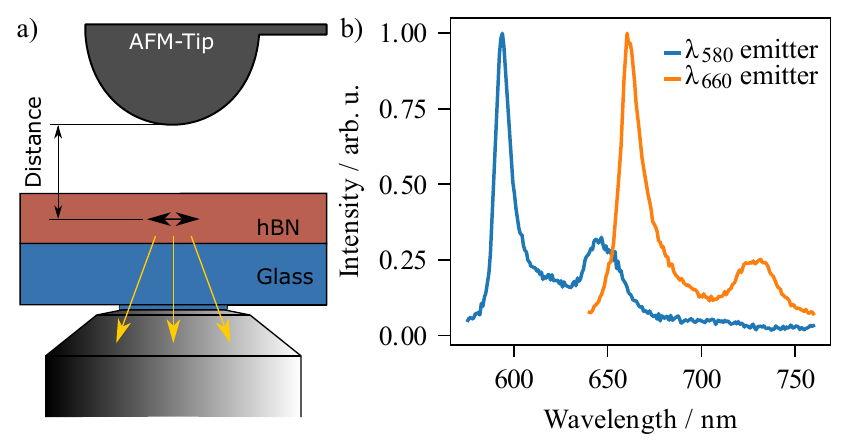}%
 \caption{
    \textit{Schematic representation of the experiment and two representative spectra of both investigated emitter families.}
    a) An AFM equipped with a gold-coated hemispherical tip with a diameter of \SI{5.5}{\micro\meter} is aligned with a SPE in hBN and held at a distance $d$. An oil immersion objective lens excites the SPE and collects its emission from below the glass substrate. b) Representative spectra of a typical SPE from each emitter family showing a pronounced ZPL at $\approx \SI{595}{\nano\meter}$ and $\SI{660}{\nano\meter}$ and a phonon sideband, respectively.
 }
 \label{fig:schematics}
\end{figure}

\begin{figure*}[htb]
 \centering
 \includegraphics[width=\textwidth]{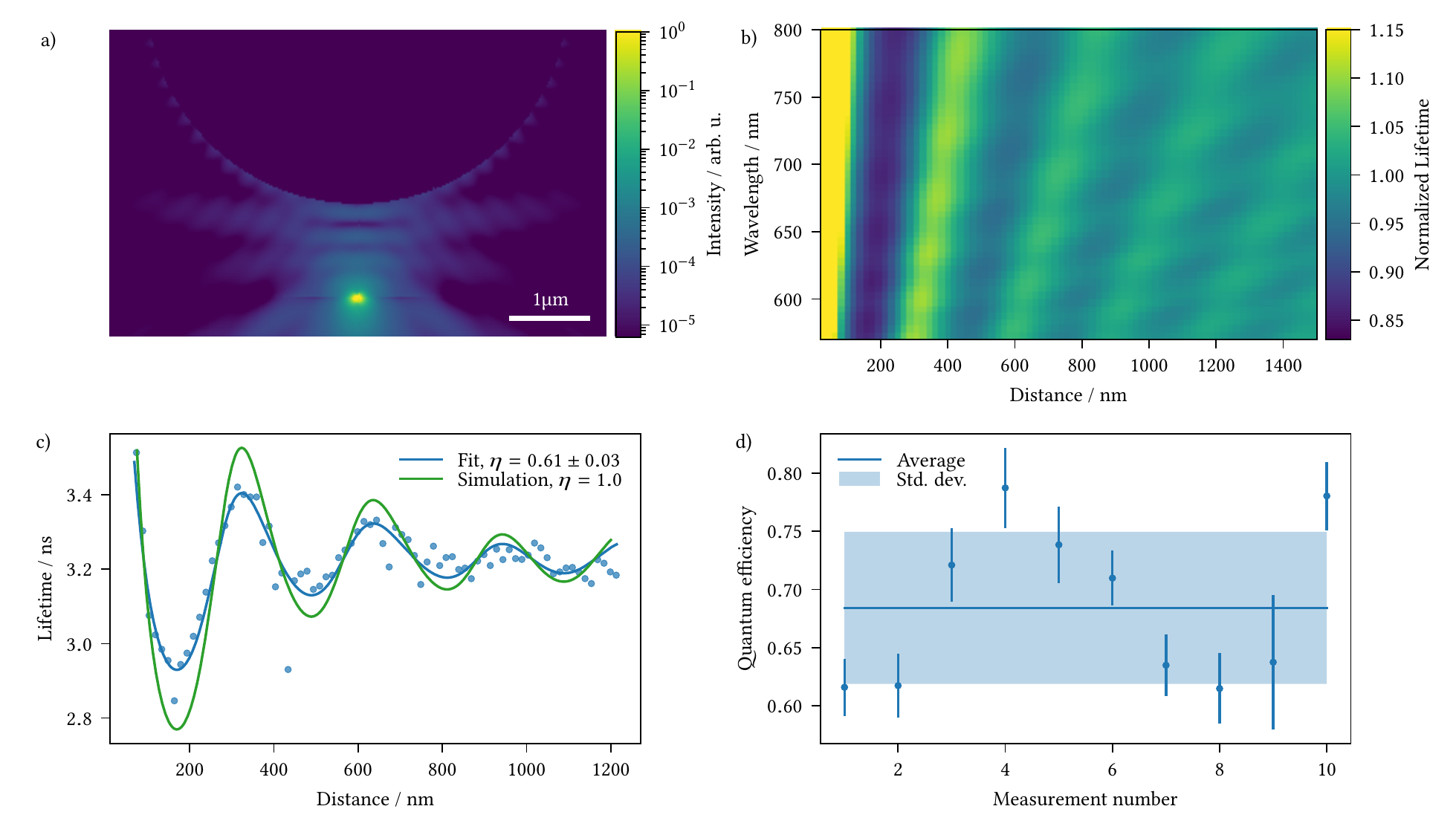}%
 \caption{
    \textit{Simulated fluorescence intensity, simulated wavelength and distance dependent lifetimes together with measured quantum efficiency data.}
    a) A simulation of the experimental scenario showing the intensity distribution of a horizontally polarized dipole (mimicking the hBN emitter), emitting at \SI{600}{\nano\meter} placed in the center of a \SI{10}{\nano\meter} thick hBN-layer, on top of a glass substrate at a distance of \SI{1240}{\nano\meter} from the gold hemisphere. b) Changes of the lifetime as a function of the distance between the AFM tip and an SPE simulated for a range of ZPL positions. The values were extracted from simulations similar to a). This map is used to fit the experimental data. c) A typical distance-dependent lifetime measurement (dots) fitted by \autoref{eq:fit} (blue solid line) determines the QE. A function with a QE of 1.0 is shown for reference by the green solid line. d) A repeated measurement of the QE (dots) is shown, together with the average (solid line) and the standard deviation (shaded area) of these measurements.
 }
 \label{fig:qe_measurement}
\end{figure*}

\begin{figure}[htb]
 \includegraphics[width=\columnwidth]{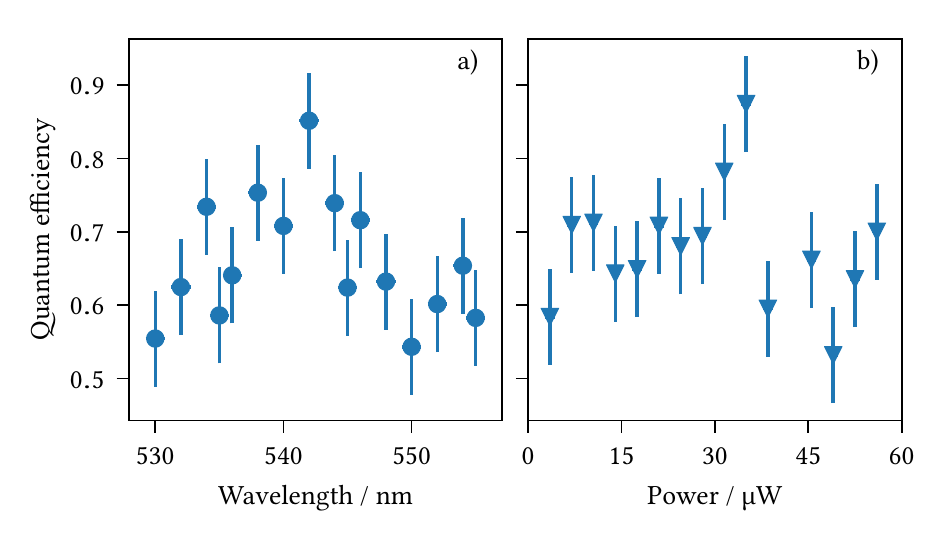}%
 \caption{
    \textit{Excitation wavelength and power dependence.}
    a) QE relative to the excitation wavelength showing a slight trend with the highest QE at \SI{540}{\nano\meter}. b) QE relative to the excitation power. Errorbars are given by the statistical error shown in \autoref{fig:qe_measurement} d).
 }
 \label{fig:energy}
\end{figure}

\begin{figure}[htb]
 \includegraphics[width=\columnwidth]{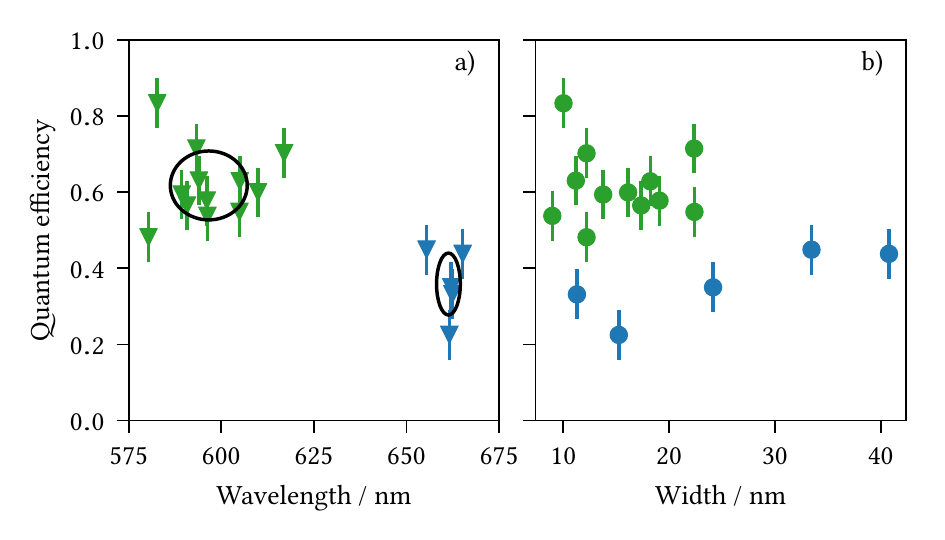}%
 \caption{
    \textit{QE dependence on ZPL width and central wavelength.}
    a) QE of 17 emitters plotted against their central ZPL position. b) The same 17 emitters plotted against their respective ZPL line width. In both plots, green marks 12 emitters with central wavelengths below \SI{640}{\nano\meter} and blue above \SI{640}{\nano\meter}. The two circles in a) are placed at the averaged position of each family, its width and height represent the respective standard deviation, given by $\lambda_{580} = \SI{597(11)}{\nano\meter}$ and $\lambda_{660} = \SI{661(4)}{\nano\meter}$. Errorbars are given by the statistical error shown in \autoref{fig:qe_measurement} d).
 }
 \label{fig:correlation}
\end{figure}

\section{Experimental setup}

The schematics of the measurement setup is shown in \autoref{fig:schematics} a). A wavelength tunable pulsed laser (Solea, Picoquant) is focused through the cover slide onto the hBN flake with an oil immersion objective lens (NA 1.4, 100x). The SPE fluorescence is collected via the same objective lens and guided through a confocal setup into a Hanbury-Brown and Twiss setup, consisting of a polarizing beam splitter (PBS) and two avalanche photon diodes (APD) (Excelitas). With a $\lambda/2$-plate before the PBS we analyzed the fluorescence polarization, finding it to be parallel to the glass substrate, an example of which is shown in the suplemental material ????. The SPEs in hBN were grown using a CVD method as described elsewhere \cite{Mendelson2019}. The hBN flakes were then transferred onto transparent glass substrate, to enable simultaneous atomic force microscope (AFM) and photo luminescence (PL) measurements. Two examples of single emitters in the hBN flakes are shown in \autoref{fig:schematics} b. To measure the QE of the SPEs, modification of their LDOS was achieved by employing a hemispherical tip with a radius of $r=\SI{2.75}{\micro\meter}$ covered by gold. Modifications of the emitter's lifetime were obtained by changing the distance of the AFM tip to the emitter. The lateral position of the tip and focus of the excitation laser were matched by scanning the tip over a large area while recording laser light reflected by the tip. Once matched, the mirrors' vertical position could be changed precisely via the build-in AFM piezo, allowing distance-dependent measurements of the SPE properties.

The QE can be considered as a scaling factor between a change in LDOS and a change in emission rate, as non-radiative processes are unaffected by a changed LDOS. A detailed discussion can be found in Ref. \cite{Drexhage1970a}. The equation used to relate QE and LDOS is given by:
\begin{equation}
    \tau(d) = \frac{\tau(\infty)}{1+\eta\left(\frac{\rho(d)}{\rho(\infty)}-1\right)}
\end{equation}
With the distance-dependent lifetimes $\tau(d)$, LDOS $\rho(d)$ and QE $\eta$, analogue to Ref. \cite{VanDam2018, Amos1997}. Accordingly, a change in LDOS is mediated by changing the mirror distance, which finally reveals the QE.

\section{Simulation of expected lifetime changes}

In order to relate the change in distance of emitter and mirror to a change in LDOS, we performed a simulation of an emitter exhibiting a horizontally polarized dipole hosted in the center of a \SI{10}{\nano\meter} thick hBN layer (refractive index of $n=1.65$) situated on top of a glass substrate. A hemisphere made from gold with a tip radius $r = \SI{2.75}{\micro\meter}$ is centered at a distance $d$ from the dipole acting as a mirror that changes the local density of states (LDOS) $\rho(d)$. The LDOS can be expressed \cite{Novotny2006} by
\begin{equation}
    \rho(d)/\rho(\infty) = p(d)/p(\infty)
\end{equation}
with the distance dependent emitted power $p$, which can be extracted from the simulation. Only SPEs with in-plane polarization were found (e.g. shown in the supplemental material ????), therefore the simulations were performed with an in-plane polarized dipole. An intensity distribution with $d=\SI{1240}{\nano\meter}$ and an emission wavelength of $\lambda=\SI{600}{\nano\meter}$ is shown in \autoref{fig:qe_measurement} a). In this situation, a standing wave pattern between dipole and mirror with three nodes can be seen. Similar simulations at different wavelength and distances were performed and relative lifetime changes extracted, results are shown in \autoref{fig:qe_measurement} b).

\section{Quantum efficiency measurement}

Once an SPE was identified, we performed distant-dependent lifetime measurements for various tip to the emitter distances. First, we fixed the AFM tip to emitter distance $d$ at approximately \SI{1.2}{\micro\meter}. At this point, we performed a lifetime measurement for \SIrange{1}{2}{\second}. Next we reduced the distance by \SI{15}{\nano\meter} and performed the next lifetime measurement, repeating this process until reaching the surface. The built-in AFM laser points at the end of the cantilever and gets reflected to a build-in four quadrant photo diode. A bending of the cantilever results in a change in position of the laser on the photo diode, which was used to indicate a completed approach to the surface and thus stopped the QE measurement. Since this method has an error margin of at least one step (\SI{15}{\nano\meter}) and we don't know the emitter's depth, we keep a distance offset $d_0$ as a fit parameter. The emission wavelength is also kept as a fit parameter within reasonable limits deducted from the measured spectrum. The maximum distance for lifetime measurements was \SI{1.2}{\micro\meter} which was sufficient to produce robust values for the lifetime at infinity $\tau_\infty$ when left as a fit parameter.


To determine the QE, we fitted the following function to the distance-dependent lifetime measurements:
\begin{equation}
    \tau(d) = \frac{\tau_\infty}{1+\eta\left(\frac{p(\lambda, d+d_0)}{p(\lambda_0, \infty)}-1\right)}.
    \label{eq:fit}
\end{equation}

\autoref{fig:qe_measurement} c) shows one representative QE measurement (dots) of a SPE with a QE of $\eta=0.61$, with a corresponding fit of \autoref{eq:fit} (blue solid line). For comparison, a case for an emitter with $\eta=1.0$ (green solid line) is also plotted.

To determine the error margins of the QE values, we performed 10 measurements on the same emitter shown by dots in \autoref{fig:qe_measurement} d). From this data set we calculated the average and the standard deviation of \SI{6.6}{\percent}, represented by the solid line and the shaded area, respectively. In the following discussion and figures, we used this standard deviation. In a reference measurement, $\tau_\infty$ changed within minutes by about one standard deviation, uncorrelated to AFM tip approaches (see supplementary material ????). Thus we speculate that photoinduced changes of the environment may cause lifetime and QE variations.

\autoref{fig:energy} a) shows the QE dependence on the excitation wavelength, conducted at an emitter with a ZPL wavelength of $\SI{595}{\nano\meter}$ emitter. In addition, a power dependent QE measurement was performed on the same emitter, using an excitation wavelength of \SI{540}{\nano\meter}. However, no systematic trend can be seen, as shown in \autoref{fig:energy} b). We therefore conclude that the QE of the hBN emitters are independent of the excitation wavelength or power, which indicates an isolated electronic structure without shelving or additional non radiative states.
Most SPEs in the CVD grown hBN sample show ZPL central positions at $\SIrange{570}{590}{\nano\meter}$ \cite{Mendelson2019} when excited with wavelength of \SI{540.0(15)}{\nano\meter}. We denote this family as $\lambda_{580}$. Consequently, we selected 12 emitters from this family at random and performed QE measurements on each of them.
Interestingly, however, when we switched the excitation laser to \SI{580(15)}{\nano\meter}, a second family of emitters shows up with ZPL central positions at $\lambda_{660} = \SI{661(4)}{\nano\meter}$, which we term $\lambda_{660}$ emitters \cite{Comtet2019}. These emitters were less common throughout the samples, and we were only able to identify five SPEs with clear ZPL (see \autoref{fig:schematics} b for example of the SPE). The $\lambda_{660}$ family may be associated with a different charge state or an altered absorption cross-section pathway. Exciting the same area of hBN with a laser wavelength in between \SI{540.0}{\nano\meter} and \SI{580.0}{\nano\meter}, didn't show any emitters with ZPL positions in between the original ZPL wavelengths.
Consequently, we compared the QE of these two families, with the results plotted in \autoref{fig:correlation} a). We find clear evidence that the emitters with the higher energy ZPL possess a higher QE. The QE of the $\eta_{580} = \SI{62(9)}{\percent}$ obtained from averaging over 12 SPEs, while the QE of the family with the longer ZPLs is $\eta_{660} = \SI{36(8)}{\percent}$, averaged over 5 emitters, respectively. We also compared the full width at half maximum (FWHM) of the emitters in both families, as shown in \autoref{fig:correlation} b). According to our results, there was no clear trend between the FWHM and the QE, for both families. This might be counter-intuitive, since it indicates that the coupling to low energy phonons does not result in non radiative transitions. Nevertheless, the clear difference in the QE values indicate that the two families have isolated electronic structures (rather than being same emitter that is shifted by strain or electric fields). 

\section{Summary}
To summarize, we presented a method to measure the absolute QE of SPEs in hBN. Accompanied by a simulation of the change in LDOS, we found record high QEs of single SPEs in hBN approaching \SI{87(7)}{\percent}. By measuring the QE of 17 SPEs and relating them to the respective ZPL wavelength, we could identify two SPE families, well separated in ZPL position, with different QEs. One family showing ZPLs clustered around \SI{580}{\nano\meter} showed an average QE of \SI{62(9)}{\percent}, while the other family operating at \SI{660}{\nano\meter} showed an average QE of \SI{36(8)}{\percent}. While the crystallographic origin of the defects is yet unknown, our results suggest that these emitters possess two distinct electronic structures.
Having ultra thin (few nm) solid state SPEs with high QEs opens up fascinating opportunities for advanced quantum photonic experiments. For instance, combining these emitters with dielectric antennas that have near unity collection efficiency, may result in a room temperature "single photon gun" \cite{Chu2017}. Such sources can also find use in quantum cryptography, that has traditionally been utilizing faint laser sources due to lack of ultra bright and ultra-pure SPEs. The SPEs in hBN that possess the higher QE can potentially meet this demand. Finally, the presented method can be extended to measure QE of localized and interlayer excitons in other 2D materials \cite{Rivera2015}. 

\section{Funding}

Financial support from the German Ministry of Education and Research (BMBF) project "NANO-FILM", the Australian Research council (via DP180100077, DP190101058), the Asian Office of Aerospace Research and Development grant FA2386-17-1-4064, the Office of Naval Research Global under grant number N62909-18-1-2025 are gratefully acknowledged. I.A. is grateful for the Humboldt Foundation for their generous support. O.B. acknowledges the UTS Distinguished Visiting Scholars scheme.

\section{Acknowledgments}

N. N. thanks Bastian Leykauf and Carlo Bradac for fruitful discussions


\bibliography{references}

\end{document}


\preprint{APS/123-QED}

\title{
Direct Measurement of Quantum Efficiency of Single Photon Emitters in Hexagonal Boron Nitride \\ Supplemental Material
}
\author{Niko Nikolay$^{1,2}$}
\author{Noah Mendelson$^3$}
\author{Ersan Özelci$^{1,2}$}
\author{Bernd Sontheimer$^{1,2}$}
\author{Florian Böhm$^{1,2}$}
\author{Günter Kewes$^{1,2}$}
\author{Milos Toth$^3$}
\author{Igor Aharonovich$^3$}
\author{Oliver Benson$^{1,2}$}
\affiliation{
 $^1$ AG Nanooptik, Humboldt-Universität zu Berlin,
 Newtonstraße 15, D-12489 Berlin, Germany \\
 $^2$ IRIS Adlershof, Humboldt-Universität zu Berlin,
 Zum Großen Windkanal 6, 12489 Berlin, Germany\\
 $^3$ School of Mathematical and Physical Sciences, University of Technology Sydney, Ultimo, New South Wales 2007, Australia
}
\date{\today}

\maketitle

\section{Lifetime and quantum efficiency measurement}

\begin{figure}[htb]
 \includegraphics[width=\columnwidth]{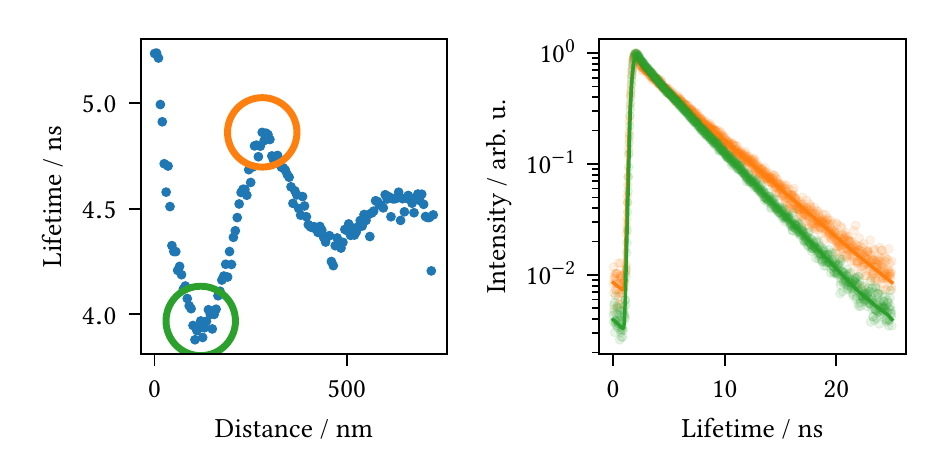}%
 \caption{
    \textit{Quantum efficiency and corresponding lifetime measurement.}
    a) Quantum efficiency measurement with 146 points and \SI{6}{\second} integration time per point, with an average count rate of \SI{200}{\kilo\countspersecond}. b) Lifetime measurement (dots) and fits (straight line), corresponding to a minimum (marked green in a) and b)) and a maximum (marked orange in a) and b)). 
 }
 \label{fig:single_emitter_lt}
\end{figure}

To determine the quantum efficiency, we related a change in local density of states (LDOS) to the changes in lifetime (see main text for detailed explanation). By approaching the single photon emitter (SPE) with a spherical mirror, a lifetime change could be observed. A typical lifetime - distance measurement can be seen in \autoref{fig:single_emitter_lt} a). Corresponding lifetime measurements and fits at the marked extreme points are shown in b). 

As described in the supplemental material of Ref. \cite{Nikolay2018}, we determined the excited state lifetime by recording time differences between the applied laser pulse and the photon arrival time at the avalanche photodiodes (APDs). A normalized histogram is shown in \autoref{fig:single_emitter_lt} b). The fit represented by the solid line is given by a circular convolution (denoted by $*$) of a double exponential decay with the instrument response function (IRF):

\begin{align}
    f(\tau) &= IRF * LT \\
    LT &= a_1 \mathrm{e}^{-t/\tau_1} + a_2 \mathrm{e}^{-t/\tau_2} + b.
\end{align}

With the amplitudes $a_1$ and $a_2$, the lifetimes $\tau_1$ and the fixed lifetime $\tau_2 = \SI{0.5}{\nano\second}$ and the offset $b$.

\section{Dipole orientation}

To reduce the free parameters needed to fit the quantum efficiency, i.e. Eq. 3 in the main text, we verified the horizontal alignment of the emitting dipole by a polarization measurement, analogous to what was described in the supplemental material of Ref. \cite{Nikolay2018}.
\begin{figure}[htb]
 \includegraphics[width=180pt]{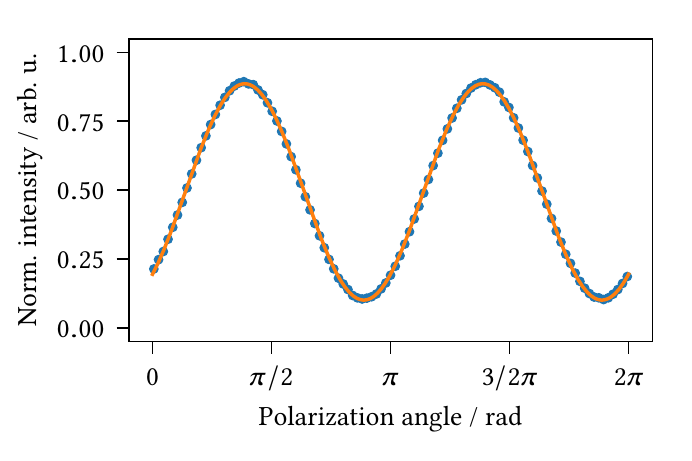}%
 \caption{
    \textit{Polarization measurement.}
    The high contrast of the shown polarization measurement of a hBN SPE indicates for a horizontally polarized dipole.
 }
 \label{fig:polarization}
\end{figure}
The fluorescence was guided through a $\lambda/2$ plate and subsequently split by a polarizing beam splitter. Each output port was directed to an avalanche photo diode (APD). The recorded intensity will be denoted as $I_\mathrm{APD_1}$ and $I_\mathrm{APD_2}$. The relative intensity detected at one port is shown in \autoref{fig:polarization} and was calculated by (additionally we normalized the relative signal to its value averaged over $2\pi$, denoted by the horizontal line):

\begin{equation}
    I_\mathrm{Norm}(\alpha) = \frac{1}{2}\,\overline{\frac{I_\mathrm{APD_1}}{I_\mathrm{APD_1}+I_\mathrm{APD_2}}}^{-1} \left(\frac{I_\mathrm{APD_1}}{I_\mathrm{APD_1}+I_\mathrm{APD_2}}\right)
\end{equation}

with the polarization angle $\alpha$. The high contrast (difference between minimal and maximal point) of \SI{.797(1)}{\arbitraryunit} indicates for a horizontally aligned dipole \cite{Lethiec2014}.

\section{Sample Preparation}
$h$-BN was fabricated \textit{via} a low pressure chemical vapor deposition process previously reported \cite{Mendelson2019}. $h$-BN was grown on a copper catalyst, using ammonia borane as a precursor. The as-grown multi-layer $h$-BN films were then transferred to a glass coverslip \textit{via} a PMMA assisted wet transfer process. The PMMA layer was then removed by soaking the sample in warm acetone overnight, before further cleaning by exposure to a UV-Ozone atmosphere for \SI{20}{\minute}.




\bibliography{references}